\documentclass[twocolumn,showpacs,preprintnumbers]{revtex4}
\usepackage{amsmath}
\usepackage{amssymb}
\usepackage{graphicx}
\usepackage{dcolumn}
\usepackage{bm}

\input{tcilatex}

\begin{document}

\preprint{APS/123-QED}
\title{Berry Phase and Adiabatic Breakdown in Optical Modulator}
\author{Eyal Buks}
\email{eyal@ee.technion.ac.il}
\affiliation{Department of Electrical Engineering, Technion, Haifa 32000, Israel }
\date{\today }

\begin{abstract}
We consider an \textit{all in-fiber} optical modulator based on a ring
resonator configuration. \ The case of adiabatic to nonadiabatic transition
is considered, where the geometrical (Berry) phase acquired in a round trip
along the ring changes abruptly by $\pi $. \ Degradation of the responsivity
of the modulator due to finite linewidth of the optical input is discussed.
\ We show that the responsivity of the proposed modulator can be
significantly enhanced with optimum design and compare with other
configurations.
\end{abstract}

\maketitle





\section{Introduction}

Optical modulators are devices of great importance for optical communication
and other fields. \ In these devices some external perturbation, e.g.
electric or magnetic fields, is employed to modulate the transmission $%
\mathcal{T}$ between the input and output optical ports. \ One of the key
property of an optical modulator is the responsivity, namely the dependence
of $\mathcal{T}$ on the applied perturbation. \ Enhancing the responsivity
is highly desirable in many applications. \ As is shown in Ref. \cite{Buks
04}, the linearity of optical modulators imposes in general an upper bound
on their responsivity. \ One way of achieving high responsivity is by
employing a resonator configuration with high quality factor $Q$. \ The
multiple back and forth reflections occurring in a resonator allow enhancing
the responsivity in comparison with the case of reflectionless optical path.
\ Such a ring resonator was considered recently by Yariv \cite{Yariv 00}, 
\cite{Yariv 02} and implemented experimentally \cite{Choi 01}, \cite{Menon
04}. \ It was shown that high enhancement is achieved when critical coupling
occurs, namely when the power entering the resonator from the input port
equals the output dissipation power. \ On the other hand, one of the
drawbacks of a resonator configuration is the limited optical bandwidth. \
In some cases finite linewidth of the optical input $\Delta \omega $ may
lead to broadening of the resonance and thus reducing the responsivity. \
Such broadening can be avoided only when $\Delta \omega /\omega <<\lambda
/QL $, where $\lambda $ is the wavelength and $L$ is a characteristic length
of the resonator.

In this paper we consider a ring resonator similar to the one discussed in 
\cite{Yariv 00}, \cite{Yariv 02}. \ However while Ref. \cite{Yariv 00}, \cite%
{Yariv 02} considered the case of polarization independent evolution, here
we study the case of finite birefringence $\mathbf{\kappa }\left( s\right) $
along the optical path ($s$ is a coordinate along the optical path). \ We
first consider the case of adiabatic evolution, when $\mathbf{\kappa }$
changes slowly. \ In this case it is convenient to express the state of
polarization (SOP) in the basis of local eigenvectors. \ In this basis the
equations of motion of both polarization amplitudes can be decoupled to the
lowest order in the adiabatic expansion. \ Next we consider the case of
adiabatic breakdown, namely the transition into the regime where the
adiabatic approximation does not hold. \ In this case the geometrical
(Berry) \cite{Berry 84} phase acquired in a round trip along the ring
changes abruptly by $\pi $. \ We show that this abrupt change can be
employed for achieving high responsivity. \ Note that similar adiabatic
breakdown was considered in Ref. \cite{Lynda-Geller 93} for the case of spin
1/2 electrons in coherent mesoscopic conductors with spin-orbit interaction
(see also Ref. \cite{Bhandari 91}).

Such an optical modulator based on adiabatic breakdown can be implemented in
a variety of different configurations. \ Here we demonstrate these effects
by considering a relatively simple example of a modulator based on a fiber
ring resonator having both intrinsic and externally applied birefringence. \
The intrinsic birefringence along the ring in our example is linear. \ As we
discuss below, it can be induced using a standard polarization maintaining
fiber being twisting and tapered to realized the desired birefringence. \
The externally applied birefringence used for modulation is based in our
example on magneto-optic effect \cite{Simon 77}, \cite{Ulrich 79}. \ This
effect allows inducing circular birefringence in the fiber, being
proportional to the Verdet constant characterizing the material and to the
component of the applied magnetic field along the direction of propagation.
\ We employ both analytical and numerical calculations to study the
responsivity of the system. \ We find enhanced responsivity when operating
in the adiabatic breakdown regime.

\section{Fiber Ring Resonator}

Consider a fiber ring resonator as seen in Fig. \ref{ring resonator}. \ It
consists of a fiber ring coupled to input and output ports using a
directional coupler.

\FRAME{ftbpFU}{2.9654in}{2.5901in}{0pt}{\Qcb{The fiber ring resonator.}}{%
\Qlb{ring resonator}}{fig1.gif}{\special{language "Scientific Word";type
"GRAPHIC";maintain-aspect-ratio TRUE;display "USEDEF";valid_file "F";width
2.9654in;height 2.5901in;depth 0pt;original-width 3.9167in;original-height
3.4169in;cropleft "0";croptop "1";cropright "1";cropbottom "0";filename
'Fig1.gif';file-properties "XNPEU";}}

The SOP at each point along the fiber is described as a \textit{spinor} with
two components associated with the amplitudes of two orthonormal
polarization states. \ As is discussed in appendix A and appendix B, we use
the local eigenvectors as a basis to express the SOP. \ The associated
amplitudes are $E_{\uparrow }$ and $E_{\downarrow }$ respectively. \ The
directional coupler is assumed to have coupling constants independent of the
SOP. \ Moreover, the coupling is assumed lossless, thus the coupling matrix
is unitary

\begin{equation}
\left( 
\begin{array}{c}
E_{_{\uparrow }}^{b_{1}} \\ 
E_{_{\uparrow }}^{b_{2}} \\ 
E_{_{\downarrow }}^{b_{1}} \\ 
E_{_{\downarrow }}^{b_{2}}%
\end{array}%
\right) =\left( 
\begin{array}{cccc}
t & r & 0 & 0 \\ 
-r^{\ast } & t^{\ast } & 0 & 0 \\ 
0 & 0 & t & r \\ 
0 & 0 & -r^{\ast } & t^{\ast }%
\end{array}%
\right) \left( 
\begin{array}{c}
E_{_{\uparrow }}^{a_{1}} \\ 
E_{_{\uparrow }}^{a_{2}} \\ 
E_{_{\downarrow }}^{a_{1}} \\ 
E_{_{\downarrow }}^{a_{2}}%
\end{array}%
\right) ,  \label{coupler}
\end{equation}

where

\begin{equation}
\left\vert t\right\vert ^{2}+\left\vert r\right\vert ^{2}=1.
\label{|r|^2+|t|^2=1}
\end{equation}

Integrating the equation of motion along the ring leads in general to a
linear relation between the amplitudes at both ends

\begin{equation}
\left( 
\begin{array}{c}
E_{_{\uparrow }}^{a_{2}} \\ 
E_{_{\downarrow }}^{a_{2}}%
\end{array}%
\right) =\hat{M}\left( 
\begin{array}{c}
E_{_{\uparrow }}^{b_{2}} \\ 
E_{_{\downarrow }}^{b_{2}}%
\end{array}%
\right) ,  \label{loop}
\end{equation}

where

\begin{equation}
\hat{M}=\left( 
\begin{array}{cc}
M_{11} & M_{12} \\ 
M_{21} & M_{22}%
\end{array}%
\right) .
\end{equation}

Using \ref{coupler}, \ref{loop}, and \ref{|r|^2+|t|^2=1} one can find a
linear relation between the amplitudes in the input and output ports of the
modulator

\begin{equation}
\left( 
\begin{array}{c}
E_{_{\uparrow }}^{b_{1}} \\ 
E_{_{\downarrow }}^{b_{1}}%
\end{array}%
\right) =\hat{S}\left( 
\begin{array}{c}
E_{_{\uparrow }}^{a_{1}} \\ 
E_{_{\downarrow }}^{a_{1}}%
\end{array}%
\right) ,
\end{equation}

where the matrix $\hat{S}$ is given by

\begin{equation}
\hat{S}=\frac{1-t\hat{M}^{-1}}{t^{\ast }-\hat{M}^{-1}}.
\end{equation}

Note that if $\hat{M}$ is unitary (namely, $\hat{M}^{-1}=\hat{M}^{\dag }$)
and \ref{|r|^2+|t|^2=1} holds then, as expected, $\hat{S}$ is unitary as
well. \ Note also that if $\hat{M}$ is diagonal (namely, $M_{12}=M_{21}=0$)
the following holds \cite{Yariv 00}, \cite{Yariv 02}

\begin{equation}
\hat{S}=\left( 
\begin{array}{cc}
\frac{t-M_{11}}{1-M_{11}t^{\ast }} & 0 \\ 
0 & \frac{t-M_{22}}{1-M_{22}t^{\ast }}%
\end{array}%
\right) .  \label{diagonal S}
\end{equation}

To find the matrix $\hat{M}$ one has to integrate the equation of motion \ref%
{de/ds} along the close curve defined by the ring. \ In the adiabatic limit,
to be discussed in the next section, the solution can be found analytically.
\ In the following section the case of adiabatic breakdown is discussed,
where both analytical approximations and numerical calculations are employed
to integrate the equation of motion \ref{de/ds}.

\section{The Adiabatic Case}

In the case where the adiabatic approximation can be applied the matrix $%
\hat{M}$ is given by

\begin{equation}
\hat{M}=\left( 
\begin{array}{cc}
\exp \left( i\delta _{\uparrow }\right) & 0 \\ 
0 & \exp \left( i\delta _{\downarrow }\right)%
\end{array}%
\right) ,
\end{equation}

where $\delta _{\uparrow }$ and $\delta _{\downarrow }$ are given by
equations \ref{delta up} and \ref{delta down} respectively.

In the more general case the ring may have internal loss. \ Assuming the
loss is polarization independent, one has

\begin{equation}
\hat{M}=\left( 1-\xi _{l}\right) \left( 
\begin{array}{cc}
\exp \left( i\delta _{\uparrow }\right) & 0 \\ 
0 & \exp \left( i\delta _{\downarrow }\right)%
\end{array}%
\right) ,
\end{equation}

where $0\leq \xi _{l}\leq 1$ is real. \ Thus using \ref{diagonal S}

\begin{equation}
\frac{E_{\sigma }^{b_{1}}}{E_{_{\sigma }}^{a_{1}}}=\frac{t-\left( 1-\xi
_{l}\right) \exp \left( i\delta _{\sigma }\right) }{1-\left( 1-\xi
_{l}\right) \exp \left( i\delta _{\sigma }\right) t^{\ast }},
\end{equation}

where $\sigma \in \left\{ \uparrow ,\downarrow \right\} $. \ Using the
notation $t=\left( 1-\xi _{c}\right) \exp \left( i\theta _{t}\right) $,
where $0\leq \xi _{c}\leq 1$ is real, and $\vartheta =\delta _{\sigma
}-\theta _{t}$ one gets

\begin{equation}
\frac{E_{\sigma }^{b_{1}}}{E_{_{\sigma }}^{a_{1}}}=\exp \left( i\theta
_{t}\right) \frac{1-\xi _{c}-\left( 1-\xi _{l}\right) \exp \left( i\vartheta
\right) }{1-\left( 1-\xi _{l}\right) \left( 1-\xi _{c}\right) \exp \left(
i\vartheta \right) }.
\end{equation}

Near resonance $\vartheta <<1$. \ Moreover, assuming $\xi _{l}<<1$ and $\xi
_{c}<<1$, one finds

\begin{equation}
\frac{E_{\sigma }^{b_{1}}}{E_{_{\sigma }}^{a_{1}}}\simeq \exp \left( i\theta
_{t}\right) \frac{\xi _{l}-\xi _{c}-i\vartheta }{\xi _{l}+\xi
_{c}-i\vartheta }.  \label{t(theta)}
\end{equation}

Critical coupling occurs when $\xi _{l}=\xi _{c}\equiv \xi $. \ In this case
the transmission amplitude $E_{\sigma }^{b_{1}}/E_{_{\sigma }}^{a_{1}}$
vanishes at resonance. \ The transmission probability in this case is given
by

\begin{equation}
\mathcal{T}\left( \vartheta \right) \simeq \frac{\left( Q\vartheta \right)
^{2}}{1+\left( Q\vartheta \right) ^{2}},  \label{T(theta)}
\end{equation}

where $Q=1/2\xi $. \ Thus, high responsivity can be achieved when operating
close to a resonance with high $Q$ factor.

\section{Broadening due to Finite Linewidth}

As was discussed in the previous section, relatively high responsivity can
be achieved when operating close to a resonance. \ However, as we discuss
below, the price one has to pay for that is limited bandwidth.

Consider the case where the optical input has some finite linewidth $\Delta
\omega $. \ As a result the phase factor $\vartheta $ will acquire a
linewidth given by

\begin{equation}
\Delta \vartheta =2\pi \frac{\Delta \omega }{\omega }\frac{L}{\lambda }.
\end{equation}

Consider the case of a polychromatic optical input and assume that the
probability distribution of $\vartheta $ is Lorenzian with a characteristic
width $\Delta \vartheta $

\begin{equation}
f\left( \vartheta ^{\prime }\right) =\frac{1}{\pi \Delta \vartheta }\frac{1}{%
1+\left( \frac{\vartheta ^{\prime }-\vartheta }{\Delta \vartheta }\right)
^{2}}.
\end{equation}

Averaging using this distribution and Eq. \ref{T(theta)}, and employing the
residue theorem for evaluating the integral one finds

\begin{eqnarray}
\mathcal{\bar{T}}\left( \vartheta \right) &=&\tint\limits_{-\infty }^{\infty
}d\vartheta ^{\prime }f\left( \vartheta ^{\prime }\right) \mathcal{T}\left(
\vartheta ^{\prime }\right) \\
&=&1-\frac{1}{1+Q\Delta \vartheta }\frac{1}{1+\left( \frac{Q\vartheta }{%
1+Q\Delta \vartheta }\right) ^{2}}.  \notag
\end{eqnarray}

Thus, for this case broadening can be avoided only if $Q\Delta \vartheta <<1$
or $\Delta \omega /\omega <<\lambda /QL$.

\section{Adiabatic Breakdown}

While in the previous case both adiabatic SOP are effectively decoupled, we
consider now the transition between adiabatic and non-adiabatic regimes.

The birefringence along the fiber ring is described by the vector $\mathbf{%
\kappa }\left( s\right) $ (see appendix A). \ Consider the case where in
some section of the ring $\mathbf{\kappa }\left( s\right) $ is close to the
degeneracy point at the origin $\mathbf{\kappa }=0$. \ In this case small
perturbation applied to $\mathbf{\kappa }\left( s\right) $ can result in a
large change in the geometrical phase \ref{gamma up} and \ref{gamma down}. \
This can be seen by considering, for example, the case of a planar curve $%
\mathbf{\kappa }\left( s\right) $. \ In this case the solid angle is given
by $\Omega =2\pi n$, where $n$ is the winding number of the curve $\mathbf{%
\kappa }\left( s\right) $ around the origin. \ As the curve $\mathbf{\kappa }%
\left( s\right) $ crosses the origin at some point, $n$ changes abruptly by
one, leading thus to an abrupt change in the geometrical phase. \ Note
however that near this transition when $\left\vert \mathbf{\kappa }\left(
s\right) \right\vert $ is small the adiabatic approximation breaks down and
alternative approaches are needed.

As an example for such a transition we consider a ring resonator for which
the close curve $\mathbf{\kappa }\left( s\right) $ has the shape seen in
Fig. \ref{kappa and P} (c) in the unperturbed case. \ This curve is made of
'half circle' section in the 1-3 plane (the linear birefringence plane) and
a 'diameter' section along the $\kappa _{3}$ axis crossing the origin. \
Such a structure can be realized by using a polarization maintaining fiber
and by employing fiber tapering techniques. \ The half circle section can be
made out of a M\"{o}bius like ring made of the polarization maintaining
fiber. \ After welding the two ends of the twisted fiber to form the M\"{o}%
bius structure one can employ tapering techniques to form the 'diameter'
section.

The curve $\mathbf{\kappa }\left( s\right) $ is perturbed by applying a
magnetic field on part of the 'diameter' section of the fiber ring. \ Such a
perturbation contributes circular birefringence in the $\kappa _{2}$
direction (see Fig. \ref{kappa and P} (a) and (e). \ The relatively high
value of the Verdet constant in common optical fibers allows significant
magneto-optic effect with moderate applied magnetic fields. \ While the
adiabatic approximation totally breaks down in the unperturbed case of Fig. %
\ref{kappa and P} (c) when the curve $\mathbf{\kappa }\left( s\right) $
crosses the origin, the perturbation transforms the system into the regime
where adiabaticity holds. \ As is shown below, the responsivity of the
system is relatively high when operating near this transition between the
adiabatic and non-adiabatic regimes.

\FRAME{ftbpFU}{3.4056in}{5.7363in}{0pt}{\Qcb{The birefringence $\mathbf{%
\protect\kappa }\left( s\right) $ and polarization $\mathbf{P}\left(
s\right) $ along the fiber ring. \ Plots (c) and (d) shows the unperturbed
case, while in (a) and (b) the perturbation parameter is $\protect\alpha %
=-0.2$, and in (e) and (f) $\protect\alpha =0.2$.}}{\Qlb{kappa and P}}{%
fig2.gif}{\special{language "Scientific Word";type
"GRAPHIC";maintain-aspect-ratio TRUE;display "USEDEF";valid_file "F";width
3.4056in;height 5.7363in;depth 0pt;original-width 5.1975in;original-height
8.7813in;cropleft "0";croptop "1";cropright "1";cropbottom "0";filename
'Fig2.gif';file-properties "XNPEU";}}

The 'half circle' section is analyzed in appendix C. \ As can be seen in
Fig. \ref{p_z vs. Lambda}, the Zener transition probability $p_{z}$ vanishes
for a series of points denoted as $\Lambda _{n}$. \ In our example we chose $%
\Lambda $ to be the first zero of $p_{z}\left( \Lambda \right) $, namely $%
\Lambda =$ $\Lambda _{1}=1.022$. \ One advantage of choosing one of the
zeros of $p_{z}\left( \Lambda \right) $, where $p_{z}$ obtains a local
minimum, is the fact that $p_{z}$ is only weakly affected by small
deviations of $\mathbf{\kappa }\left( s\right) $ from the ideal 'half
circle' curve. For the parameter $\gamma $ we chose the value $\gamma =1$. \
As can be seen from Fig. \ref{kappa and P} (d) for this choice the evolution
along the 'half circle' section transform the polarization vector on the
Bloch sphere from the pole on the negative $P_{z}$ axis to the opposite pole
on the positive $P_{z}$ axis. \ The fiber length of this section is $%
2\Lambda _{1}/\gamma $.

The rest of the fiber ring has a birefringence given by $\mathbf{\kappa }%
\left( s\right) =\mathbf{\kappa }_{0}\left( s\right) +\mathbf{\kappa }%
_{1}\left( s\right) $, where $\mathbf{\kappa }_{0}\left( s\right) $ is the
unperturbed birefringence forming the 'diameter' section and $\mathbf{\kappa 
}_{1}\left( s\right) $ is the perturbation induced by the magnetic field. \
The unperturbed part is assumed to be given by

\begin{equation}
\mathbf{\kappa }_{0}\left( s\right) =-\left( 0,0,\frac{\Lambda _{1}\gamma
^{2}}{\beta }s\right) ,
\end{equation}

where $\left\vert s\right\vert <\beta /\gamma $. \ In our numerical example
the dimensionless parameter $\beta $ is given the value $\beta =5$. \ The
perturbation due to the applied magnetic field gives rise to birefringence
given by

\begin{equation}
\mathbf{\kappa }_{1}\left( s\right) =\left( 0,\frac{\alpha }{1+\exp A\left[
\left( \frac{\gamma s}{\beta }\right) ^{2}-B^{2}\right] },0\right) ,
\end{equation}

where $A=50$ and $B=0.6$ in our numerical example. \ Thus, the magnetic
field is applied to a fiber section of length $2B\beta /\gamma $ and drops
down to zero abruptly outside this section (due to the large value chosen
for the parameter $A$). \ The coupling constants in the numerical example
are $\xi _{c}=10^{-2}$ and $\xi _{l}=10^{-4}$.

The equation of motion along the fiber ring is integrated numerically as
described in appendix A. \ This allows calculating the evolution of the
polarization vector on the Bloch sphere (see Fig. \ref{kappa and P} (b),
(d), and (f)). \ The same calculation yields also the matrix $\hat{M}$. \
The off-diagonal matrix elements allow calculating the Zener transition
probability $\left\vert M_{12}\right\vert ^{2}=\left\vert M_{21}\right\vert
^{2}$ (see Fig. \ref{vs. alpha} (a) solid line). \ The curve shows the
gradual transition between the non-adiabatic limit where $\left\vert \alpha
\right\vert <<1$ and the adiabatic limit $\left\vert \alpha \right\vert >>1$%
. \ An approximated analytical expression for the Zener probability in a
similar case where the curve $\mathbf{\kappa }\left( s\right) $ is an
infinite straight line was derived in appendix C. \ The result in Eq. \ref%
{p_z staright line} can be used to estimate approximately the Zener
transition probability for the present example

\begin{equation}
p_{z}=\exp \left( -\frac{\pi \beta \alpha ^{2}}{\Lambda _{1}\gamma ^{2}}%
\right) .  \label{p_z ana}
\end{equation}

The estimate in Eq. \ref{p_z ana} is shown in Fig. \ref{vs. alpha} (a) as a
dashed line. \ The deviation between the numerical and analytical results is
originated mainly by the fact that the straight line section in $\mathbf{%
\kappa }\left( s\right) $ is finite while the analytical analysis assumes an
infinite straight line. \ Moreover, the analytical result is expected to
hold only in the limit where $\left\vert p_{z}\right\vert <<1$ as it is
evaluated only to lowest order in the adiabatic expansion.

Figure \ref{vs. alpha} (b) shows the phase of both diagonal matrix elements
of $\hat{M}$. \ In both cases the phase changes abruptly by $\pi $ near $%
\alpha =0$. \ This is originated by the sharp change of the solid angle $%
\Omega $ by $2\pi $ near $\alpha =0$ (see Eq. \ref{delta up} and \ref{delta
down}). \ The optical modulator discussed in the present work employs this
sharp change to achieve high responsivity.

Figure \ref{vs. alpha} (c) shows the transmission probability into both SOP, 
$P_{11}=\left\vert S_{11}\right\vert ^{2}$ (solid line) and $%
P_{21}=\left\vert S_{21}\right\vert ^{2}$ (dashed line) of the entire
modulator. \ For both cases, the full width half maximum (FWHM) is $\Delta
\alpha =5.1\times 10^{-3}$.

\FRAME{ftbpFU}{3.2673in}{2.3402in}{0pt}{\Qcb{Dependence on the perturbation
amplitude $\protect\alpha $. \ (a) Zener transition probability, calculated
numerically (solid line) and estimated using Eq. \protect\ref{p_z staright
line}. \ (b) \ Phase of $M_{11}$ (solid line) and of $M_{22}$ (dashed line).
\ (c) \ transmission probability into both SOP, $P_{11}=\left\vert
S_{11}\right\vert ^{2}$ (solid line) and $P_{21}=\left\vert
S_{21}\right\vert ^{2}$ (dashed line).}}{\Qlb{vs. alpha}}{fig3.gif}{\special%
{language "Scientific Word";type "GRAPHIC";maintain-aspect-ratio
TRUE;display "USEDEF";valid_file "F";width 3.2673in;height 2.3402in;depth
0pt;original-width 6.4792in;original-height 4.625in;cropleft "0";croptop
"1";cropright "1";cropbottom "0";filename 'Fig3.gif';file-properties
"XNPEU";}}

\section{Discussion}

As we have seen, the ring resonator can serve as an optical modulator with
high responsivity when operated near one of its resonances. \ Two regimes of
operation were considered, the adiabatic one, and the non-adiabatic one. \
In what follows we compare between both regimes by considering the following
points.

\textbf{Optical Source Linewidth - }In the adiabatic limit, when the
equations of motion in the adiabatic basis become decoupled, the only effect
of the external perturbation is on the phases acquired along the fiber ring.
\ The dependence of the dynamical phase on wavelength gives rise to
broadening of resonances when operating with an optical input having finite
linewidth. \ In the general nonadiabatic regime, however, the external
perturbation can affect not only the phase factors but also the SOP as it
evolves along the close fiber ring. \ The later, being wavelength
independent gives rise to a modified dependence on the optical source
linewidth.

\textbf{Critical Coupling} - In the adiabatic regime full modulation between
zero and one of the transmission probability $\mathcal{T}$ is possible only
when critical coupling occurs, namely $\xi _{c}=\xi _{l}$, (see Eq. \ref%
{t(theta)}). \ In practice, fulfilling this condition when $\xi _{c}=\xi
_{l}<<1$ is difficult. \ However, this condition is not essential in the
general non-adiabatic case. \ As can be seen in Fig. \ref{vs. alpha} (c)
full modulation is achieved, even thought for this example $\xi _{c}=100\xi
_{l}$.

\textbf{Responsivity} - The responsivity of the ring resonator device can be
characterized by the FWHM and height of the resonance near which the device
is being operated. \ As was discussed above, $\Delta \alpha =5.1\times
10^{-3}$ for the example presented in Fig. \ref{vs. alpha}. \ For the same
parameters the FWHM of the resonances in the adiabatic regime $\left\vert
\alpha \right\vert >>1$ can be evaluated using Eq. \ref{t(theta)} yielding $%
\Delta \alpha =1.7\times 10^{-3}$. \ However, as was discussed above, since
the coupling is not critical, the modulation is not full in the adiabatic
case. \ Note that in general the responsivity has an upper bound imposed by
the linearity of the system \cite{Buks 04}. \ It can be shown that for both
cases, the obtained responsivity is of the same order as the upper bound. \
A future publication will discuss this point in more details.

\section{Summary}

In the present work we study an optical modulator based on a fiber ring
resonator. \ Both adiabatic and non-adiabatic regimes of operation are
considered. \ We find that operating close to the point where the
geometrical phase changes abruptly by $\pi $ can allow relatively high
responsivity, even when coupling is not set to be critical. \ Our example
deals with a particular configuration of optical fiber ring with both
intrinsic and externally applied birefringence. \ However, the same ideas
can be implemented with other optical waveguides and other birefringence
mechanisms.

\section{Acknowledgements}

The author would especially like to thank Avishai Eyal for many helpful
conversations and invaluable suggestions. \ Also a discussion with Steve
Lipson is gratefully acknowledged.

\appendix

\section{SOP Evolution Along a Fiber}

Consider an optical fiber winded in some spacial curve in space. \ Let $%
\mathbf{r}\left( s\right) $ be an arc-length parametrization of this curve,
namely the tangent $\mathbf{\hat{s}=}d\mathbf{r}/ds$ is a unit vector. \ The
normal unit vector $\mathbf{\hat{\nu}}$ and the curvature $\kappa $ are
defined as $d\mathbf{\hat{s}/}ds=\kappa \mathbf{\hat{\nu}}$. \ One can
easily show that $\mathbf{\hat{\nu}}\cdot \mathbf{\hat{s}}=0$ by taking the
derivative of $\mathbf{\hat{s}\cdot \hat{s}}=1$ with respect to $s$. \ The
vectors $\mathbf{\hat{s}}$, $\mathbf{\hat{\nu}}$ and the binormal unit
vector, defined as $\mathbf{\hat{b}}=\mathbf{\hat{s}\times \hat{\nu}}$, form
a local triplet orthonormal coordinate frame known as Serret - Frenet frame 
\cite{Ross 84}, \cite{Tomita 86} (see Fig. \ref{Seret-Frenet frame}). \ By
taking the derivative of $\mathbf{\hat{s}\cdot \hat{\nu}}=0$ with respect to 
$s$ one finds $\mathbf{\hat{s}\cdot }d\mathbf{\hat{\nu}/}ds=-\kappa $. \
Similarly, by taking the derivative of $\mathbf{\hat{b}}\cdot \mathbf{\hat{%
\nu}}=0$ with respect to $s$ one finds $\mathbf{\hat{b}}\cdot d\mathbf{\hat{%
\nu}/}ds=-\mathbf{\hat{\nu}}\cdot d\mathbf{\hat{b}/}ds$. \ Using the
definition $\mathbf{\hat{b}}=\mathbf{\hat{s}}\times \mathbf{\hat{\nu}}$ one
finds $d\mathbf{\hat{b}/}ds=\mathbf{\mathbf{\hat{s}}}\times d\mathbf{\hat{\nu%
}/}ds$. \ Thus $\mathbf{\hat{s}\cdot }d\mathbf{\hat{b}/}ds=0$. \ Moreover,
by taking the derivative of $\mathbf{\hat{b}}\cdot \mathbf{\hat{b}}=1$ with
respect to $s$ on finds $\mathbf{\hat{b}\cdot }d\mathbf{\hat{b}/}ds=0$. \
Thus $d\mathbf{\hat{b}}/ds$ is parallel to $\mathbf{\hat{\nu}}$. \ The
torsion $\tau $ is defined as $d\mathbf{\hat{b}/}ds=-\tau \mathbf{\hat{\nu}}$%
. \ The above definitions and relations can be summarized as follows

\begin{equation}
\frac{d}{ds}\left( 
\begin{array}{c}
\mathbf{\hat{s}} \\ 
\mathbf{\hat{\nu}} \\ 
\mathbf{\hat{b}}%
\end{array}%
\right) =\left( 
\begin{array}{ccc}
0 & \kappa & 0 \\ 
-\kappa & 0 & \tau \\ 
0 & -\tau & 0%
\end{array}%
\right) \left( 
\begin{array}{c}
\mathbf{\hat{s}} \\ 
\mathbf{\hat{\nu}} \\ 
\mathbf{\hat{b}}%
\end{array}%
\right) .  \label{d/ds}
\end{equation}

\FRAME{ftbpFU}{2.9118in}{2.8106in}{0pt}{\Qcb{The Serret-Frenet frame.}}{\Qlb{%
Seret-Frenet frame}}{fig4.gif}{\special{language "Scientific Word";type
"GRAPHIC";maintain-aspect-ratio TRUE;display "USEDEF";valid_file "F";width
2.9118in;height 2.8106in;depth 0pt;original-width 3.6045in;original-height
3.4791in;cropleft "0";croptop "1";cropright "1";cropbottom "0";filename
'Fig4.gif';file-properties "XNPEU";}}

The equation of motion along the optical ray defined by the fiber can be
obtained using the transport equation of geometric optics \cite{Kravtsov}
for the electric field phasor $\mathbf{E}_{0}$

\begin{gather}
2\left( \nabla \psi \cdot \nabla \right) \mathbf{E}_{0}+\mathbf{E}_{0}\left[
\nabla ^{2}\psi -\nabla \left( \ln \mu \right) \cdot \nabla \psi \right] \\
+2\left[ \mathbf{E}_{0}\cdot \nabla \left( \ln n\right) \right] \nabla \psi
=0,  \notag
\end{gather}

where $\psi $ is the eikonal, $n$ is the index of refraction, and $\mu $ is
the permeability. \ Define the unit vector $\mathbf{\hat{e}}_{0}=\mathbf{E}%
_{0}/\sqrt{\mathbf{E}_{0}\cdot \mathbf{E}_{0}^{\ast }}$ in the direction of $%
\mathbf{E}_{0}$. \ In terms of $\mathbf{\hat{e}}_{0}$ the transport equation
reads

\begin{equation}
\frac{d}{ds}\mathbf{\hat{e}}_{0}=-\kappa \left( \mathbf{\hat{e}}_{0}\cdot 
\mathbf{\hat{\nu}}\right) \mathbf{\hat{s}.}
\end{equation}

Expressing the unit vector $\mathbf{\hat{e}}_{0}$ in the Serret - Frenet
frame

\begin{equation}
\mathbf{\hat{e}}_{0}=e_{\nu }\mathbf{\hat{\nu}}+e_{b}\mathbf{\hat{b},}
\end{equation}

one finds using \ref{d/ds}

\begin{equation}
\frac{de_{\nu }}{ds}\mathbf{\hat{\nu}+}\frac{de_{b}}{ds}\mathbf{\hat{b}}%
+e_{\nu }\left( -\kappa \mathbf{\hat{s}+}\tau \mathbf{\hat{b}}\right)
-e_{b}\tau \mathbf{\hat{\nu}}=-\kappa e_{\nu }\mathbf{\hat{s},}
\end{equation}

thus, using the Dirac \textit{ket} notation

\begin{equation}
\left\vert e\right\rangle \dot{=}\left( 
\begin{array}{c}
e_{\nu } \\ 
e_{b}%
\end{array}%
\right) \mathbf{,}
\end{equation}

one finds

\begin{equation}
\frac{d}{ds}\left\vert e\right\rangle =i\mathcal{K}_{g}\left\vert
e\right\rangle \mathbf{,}  \label{d/ds(e_ni,e_b)}
\end{equation}

where the geometrical birefringence $\mathcal{K}_{g}$ is given by

\begin{equation}
\mathcal{K}_{g}=\tau \left( 
\begin{array}{cc}
0 & -i \\ 
i & 0%
\end{array}%
\right) \mathbf{.}
\end{equation}

Equation \ref{d/ds(e_ni,e_b)} is known as Rytov's law \cite{Kravtsov}. \ In
the more general case where other birefringence mechanisms are present the
equation of motion reads

\begin{equation}
\frac{d}{ds}\left\vert e\right\rangle =i\mathcal{K}\left\vert e\right\rangle 
\mathbf{,}  \label{de/ds}
\end{equation}

where $\mathcal{K}=\mathcal{K}_{g}+\mathcal{K}_{f}$, and $\mathcal{K}_{f}$
is the birefringence in the fiber due to intrinsic structure or due to
elasto-optic or electro-optic of magneto-optic effects.

In a lossless fiber the matrix $\mathcal{K}$ is Hermitian. \ For this case
it is convenient to expresses $\mathcal{K}$ as

\begin{equation}
\mathcal{K}=k_{0}I+\mathbf{\kappa }\cdot \mathbf{\sigma },
\label{big and small kappa}
\end{equation}

where $I$ is the 2 by 2 identity matrix, $k_{0}$ is a real scalar, $\mathbf{%
\kappa =}\left( \kappa _{1},\kappa _{2},\kappa _{3}\right) $ is a
three-dimensional real vector and the components of the Pauli matrix vector $%
\mathbf{\sigma }$\ are given by:

\begin{equation}
\sigma _{1}=\left( 
\begin{array}{cc}
0 & 1 \\ 
1 & 0%
\end{array}%
\right) ,\sigma _{2}=\left( 
\begin{array}{cc}
0 & -i \\ 
i & 0%
\end{array}%
\right) ,\sigma _{3}=\left( 
\begin{array}{cc}
1 & 0 \\ 
0 & -1%
\end{array}%
\right) .
\end{equation}

The $s$ - evolution operator $u\left( s,s_{0}\right) $ of the equation of
motion \ref{de/ds} relates an initial state $\left\vert e\left( s_{0}\right)
\right\rangle $ with a final state at some $s>s_{0}$

\begin{equation}
\left\vert e\left( s\right) \right\rangle =u\left( s,s_{0}\right) \left\vert
e\left( s_{0}\right) \right\rangle .
\end{equation}

It can be expressed as

\begin{equation}
u\left( s,s_{0}\right) =\lim_{N\rightarrow \infty
}\tprod\limits_{n=1}^{N}\exp \left[ i\frac{\Delta s}{N}\mathcal{K}\left(
s_{n}\right) \right] ,  \label{u(s,s_0)}
\end{equation}

where $\Delta s=s-s_{0}$, and $s_{n}=s_{0}+n\Delta s/N$. \ For a finite $N$
the above expression can be used as a numerical approximation of $u\left(
s,s_{0}\right) $. \ For calculating the exponential terms in \ref{u(s,s_0)}
it is useful to employ the following identity

\begin{equation}
\exp \left( ix\mathcal{K}\right) =\exp \left( ik_{0}x\right) \left[ I\cos
\left( \alpha x\right) +i\mathbf{\hat{\kappa}}\cdot \mathbf{\sigma }\sin
\left( \alpha x\right) \right] ,
\end{equation}

where the notation of Eq. \ref{big and small kappa} is being used, and $%
\mathbf{\kappa }=\mathbf{\hat{\kappa}}\alpha $ where $\mathbf{\hat{\kappa}}$
is a unit vector and $\alpha =\left\vert \mathbf{\kappa }\right\vert $.

The normalized SOP $\left\vert e\right\rangle $ can be represented as a
point on the Bloch sphere indicating the expectation value of the Pauli spin
vector matrix, namely

\begin{equation}
\mathbf{P=}\left\langle e\right\vert \mathbf{\sigma }\left\vert
e\right\rangle .
\end{equation}

\section{The Adiabatic Case}

To establish notation we review below the main results of Ref. \cite{Berry
84}. \ Consider the differential equation

\begin{equation}
\frac{d}{ds}\left\vert \psi \right\rangle =i\mathcal{K}\left\vert \psi
\right\rangle ,  \label{d_ksi/ds}
\end{equation}

where $\left\vert \psi \right\rangle $ represents $N$ dimensional column
vector and $\mathcal{K}=\mathcal{K}\left( s\right) $ is $N\times N$
Hermitian matrix. \ For any given value of $s$ the Hermitian matrix $%
\mathcal{K}\left( s\right) $ has a set of orthonormal eigenvectors

\begin{equation}
\mathcal{K}\left\vert n\left( s\right) \right\rangle =K_{n}\left( s\right)
\left\vert n\left( s\right) \right\rangle ,  \label{K|n>=K_n|n>}
\end{equation}

where $n=1,2,...,N$ and

\begin{equation}
\left\langle n\left( s\right) |m\left( s\right) \right\rangle =\delta _{nm}.
\end{equation}

The solution can be expanded as follows

\begin{equation}
\left\vert \psi \right\rangle =\sum\limits_{n}a_{n}\left( s\right) \exp 
\left[ i\int_{0}^{s}ds^{\prime }K_{n}\left( s^{\prime }\right) \right]
\left\vert n\left( s\right) \right\rangle .
\end{equation}

Substituting in \ref{d_ksi/ds} yields

\begin{equation}
\dot{a}_{m}\left( s\right) =-\sum\limits_{n}a_{n}\left( s\right)
e^{i\int_{0}^{s}ds^{\prime }\left[ K_{n}\left( s^{\prime }\right)
-K_{m}\left( s^{\prime }\right) \right] }\left\langle m\left( s\right) |\dot{%
n}\left( s\right) \right\rangle ,  \label{adot_m(s)}
\end{equation}

where upper-dot represents derivative with respect to $s$. \ The
off-diagonal terms, given by

\begin{equation}
\left\langle m\left( s\right) |\dot{n}\left( s\right) \right\rangle =\frac{%
\left\langle m\left( s\right) \right\vert \mathcal{\dot{K}}\left\vert
n\left( s\right) \right\rangle }{K_{n}\left( s\right) -K_{m}},
\end{equation}

where $m\neq n$, are neglected in the adiabatic approximation. \ The
resulting decoupled set of equations are easily solved

\begin{equation}
a_{m}\left( s\right) =a_{m}\left( 0\right) \exp \left( i\gamma _{m}\right) .
\end{equation}

where the real phase $\gamma _{m}$ is given by

\begin{equation}
\gamma _{m}=i\int\limits_{0}^{s}ds^{\prime }\left\langle m\left( s^{\prime
}\right) |\dot{m}\left( s^{\prime }\right) \right\rangle .
\end{equation}

Consider now the two dimensional case $N=2$. \ Using the notation of Eq. \ref%
{big and small kappa} and the notation $\mathbf{\kappa }=\mathbf{\hat{\kappa}%
}\alpha $, where $\mathbf{\hat{\kappa}}$ is a unit vector, given in
spherical coordinates by

\begin{equation}
\mathbf{\hat{\kappa}}=\left( \cos \varphi \sin \theta ,\sin \varphi \sin
\theta ,\cos \theta \right) ,
\end{equation}

one finds

\begin{equation}
\mathcal{K}=k_{0}I+\alpha \left( 
\begin{array}{cc}
\cos \theta & \sin \theta \exp \left( -i\varphi \right) \\ 
\sin \theta \exp \left( i\varphi \right) & -\cos \theta%
\end{array}%
\right) .
\end{equation}

The orthonormal eigenvectors are chosen to be

\begin{equation}
\left\vert \uparrow \right\rangle =\left( 
\begin{array}{c}
\cos \frac{\theta }{2}\exp \left( -\frac{i\varphi }{2}\right) \\ 
\sin \frac{\theta }{2}\exp \left( \frac{i\varphi }{2}\right)%
\end{array}%
\right) ,\;\left\vert \downarrow \right\rangle =\left( 
\begin{array}{c}
-\sin \frac{\theta }{2}\exp \left( -\frac{i\varphi }{2}\right) \\ 
\cos \frac{\theta }{2}\exp \left( \frac{i\varphi }{2}\right)%
\end{array}%
\right) ,  \label{up,down}
\end{equation}

and the following holds $\left\langle \uparrow |\uparrow \right\rangle
=\left\langle \downarrow |\downarrow \right\rangle =1$, $\left\langle
\uparrow |\downarrow \right\rangle =0$, and

\begin{equation}
\mathcal{K}\left\vert \uparrow \right\rangle =\left( k_{0}+\alpha \right)
\left\vert \uparrow \right\rangle ,
\end{equation}

\begin{equation}
\mathcal{K}\left\vert \downarrow \right\rangle =\left( k_{0}-\alpha \right)
\left\vert \downarrow \right\rangle .
\end{equation}

The eigenstates $\left\vert n\right\rangle $ (where $n\in \left\{ \uparrow
,\downarrow \right\} $) are independent of $k_{0}$, thus:

\begin{equation}
\gamma _{m}=i\int\limits_{s_{1}}^{s_{2}}ds\left\langle m\left( s\right) |%
\dot{m}\left( s\right) \right\rangle =i\int\limits_{\mathbf{\kappa }_{1}}^{%
\mathbf{\kappa }_{2}}d\mathbf{\kappa }\cdot \left\langle m\left( \mathbf{%
\kappa }\right) \right\vert \nabla _{\mathbf{\kappa }}\left\vert m\left( 
\mathbf{\kappa }\right) \right\rangle .
\end{equation}

Using the expression for a gradient in spherical coordinates one finds

\begin{equation}
\left\langle \uparrow \right\vert \nabla _{\mathbf{\kappa }}\left\vert
\uparrow \right\rangle =-\frac{i\mathbf{\hat{\varphi}}}{2\alpha }\func{ctg}%
\theta ,
\end{equation}

\begin{equation}
\left\langle \downarrow \right\vert \nabla _{\mathbf{\kappa }}\left\vert
\downarrow \right\rangle =\frac{i\mathbf{\hat{\varphi}}}{2\alpha }\func{ctg}%
\theta ,
\end{equation}

For the case of a close path, Stock's theorem can be used to express the
integral in terms of a surface integral over the surface bounded by the
close curve $\mathbf{\kappa }\left( s\right) $

\begin{equation}
\gamma _{m}=i\oint d\mathbf{\kappa }\cdot \left\langle m\right\vert \nabla _{%
\mathbf{\kappa }}\left\vert m\right\rangle =i\int\limits_{S}d\mathbf{a}\cdot
\left( \nabla \times \left\langle m\right\vert \nabla _{\mathbf{\kappa }%
}\left\vert m\right\rangle \right) .
\end{equation}

Expressing the curl operator in spherical coordinates one finds

\begin{equation}
\nabla \times \left\langle \uparrow \right\vert \nabla _{\mathbf{\kappa }%
}\left\vert \uparrow \right\rangle =\frac{i}{2}\frac{\mathbf{\kappa }}{%
\left\vert \mathbf{\kappa }\right\vert ^{3}},
\end{equation}

\begin{equation}
\nabla \times \left\langle \downarrow \right\vert \nabla _{\mathbf{\kappa }%
}\left\vert \downarrow \right\rangle =-\frac{i}{2}\frac{\mathbf{\kappa }}{%
\left\vert \mathbf{\kappa }\right\vert ^{3}},
\end{equation}

and

\begin{equation}
\gamma _{\uparrow }=-\frac{1}{2}\int\limits_{S}d\mathbf{a}\cdot \frac{%
\mathbf{\kappa }}{\left\vert \mathbf{\kappa }\right\vert ^{3}}=-\frac{1}{2}%
\Omega ,  \label{gamma up}
\end{equation}

\begin{equation}
\gamma _{\downarrow }=\frac{1}{2}\int\limits_{S}d\mathbf{a}\cdot \frac{%
\mathbf{\kappa }}{\left\vert \mathbf{\kappa }\right\vert ^{3}}=\frac{1}{2}%
\Omega ,  \label{gamma down}
\end{equation}

where $\Omega $ is the solid angle subtended by the close path $\mathbf{%
\kappa }\left( s\right) $ as seen from the origin. \ Due to the geometrical
nature of the last result, the phase factors $\gamma _{\uparrow }$ and $%
\gamma _{\downarrow }$ are called geometrical phases.

Thus

\begin{equation}
\left\vert \psi \left( s\right) \right\rangle =\left( 
\begin{array}{cc}
\exp \left( i\delta _{\uparrow }\right) & 0 \\ 
0 & \exp \left( i\delta _{\downarrow }\right)%
\end{array}%
\right) \left\vert \psi \left( 0\right) \right\rangle ,
\label{ksi(s) adiabatic}
\end{equation}

where

\begin{equation}
\delta _{\uparrow }=-\frac{\Omega }{2}+\int_{0}^{s}ds^{\prime }\left[
k_{0}\left( s^{\prime }\right) +\alpha \left( s^{\prime }\right) \right] ,
\label{delta up}
\end{equation}

\begin{equation}
\delta _{\downarrow }=\frac{\Omega }{2}+\int_{0}^{s}ds^{\prime }\left[
k_{0}\left( s^{\prime }\right) -\alpha \left( s^{\prime }\right) \right] ,
\label{delta down}
\end{equation}

\section{Zener Transitions}

The set of equations \ref{adot_m(s)} for the two dimensional case $N=2$ can
be written in a matrix form as follows

\begin{eqnarray}
&&\frac{d}{ds}\left( 
\begin{array}{c}
a_{\uparrow } \\ 
a_{\downarrow }%
\end{array}%
\right)  \label{d/ds(a_up,a_down_)} \\
&=&\left( 
\begin{array}{cc}
-\left\langle \uparrow |\dot{\uparrow}\right\rangle & -\exp \left( i\beta
\right) \left\langle \uparrow |\dot{\downarrow}\right\rangle \\ 
-\exp \left( -i\beta \right) \left\langle \downarrow |\dot{\uparrow}%
\right\rangle & -\left\langle \downarrow |\dot{\downarrow}\right\rangle%
\end{array}%
\right) \left( 
\begin{array}{c}
a_{\uparrow } \\ 
a_{\downarrow }%
\end{array}%
\right) ,  \notag
\end{eqnarray}

where

\begin{eqnarray}
\beta \left( s\right) &=&\int_{0}^{s}ds^{\prime }\left[ K_{\downarrow
}\left( s^{\prime }\right) -K_{\uparrow }\left( s^{\prime }\right) \right] 
\notag \\
&=&-2\int_{0}^{s}ds^{\prime }\left\vert \mathbf{\kappa }\right\vert .
\end{eqnarray}

In the adiabatic limit the off-diagonal matrix elements are considered
negligibly small, and consequently no transitions between the adiabatic
states occur. \ To calculate the transition probability to lowest order we
consider the off diagonal elements as a perturbation \cite{Lynda-Geller 93}.
\ The solution of the unperturbed problem is given by

\begin{equation}
a_{\downarrow }\left( s\right) =a_{_{\downarrow }}\left( 0\right) \exp
\left( i\gamma _{_{\downarrow }}\right) ,
\end{equation}

\begin{equation}
a_{\uparrow }\left( s\right) =a_{\uparrow }\left( 0\right) \exp \left(
i\gamma _{\uparrow }\right) .
\end{equation}

Assuming that at some initial point $s_{0}$ the system was in the $%
\left\vert \downarrow \right\rangle $ state, we wish to calculate the
probability to find the system in the $\left\vert \uparrow \right\rangle $
state at $s>s_{0}$. \ Lowest order correction is obtained by substituting
the unperturbed solution in \ref{d/ds(a_up,a_down_)}

\begin{equation}
\frac{d}{ds}a_{\uparrow }=-a_{_{\downarrow }}\left( 0\right) \exp \left[
i\left( \beta +\gamma _{_{\downarrow }}\right) \right] \left\langle \uparrow
|\dot{\downarrow}\right\rangle ,
\end{equation}

Thus, to lowest order the transition probability is given by

\begin{equation}
p_{z}=\left\vert \int_{s_{0}}^{s}ds^{\prime }\exp \left[ i\left( \beta
+\gamma _{_{\downarrow }}\right) \right] \left\langle \uparrow |\dot{%
\downarrow}\right\rangle \right\vert ^{2}.
\end{equation}

Consider the case where $\mathbf{\kappa }=\alpha \left( \cos \varphi \sin
\theta ,\sin \varphi \sin \theta ,\cos \theta \right) $ is planar with $%
\varphi =const$. \ Using \ref{up,down}

\begin{equation}
\left\vert \dot{\downarrow}\right\rangle =-\frac{\dot{\theta}}{2}\left\vert
\uparrow \right\rangle ,
\end{equation}

thus

\begin{equation}
\left\langle \uparrow |\dot{\downarrow}\right\rangle =-\frac{\dot{\theta}}{2}%
.
\end{equation}

Similarly

\begin{equation}
\left\langle \downarrow |\dot{\downarrow}\right\rangle =0,
\end{equation}

thus $\gamma _{_{\downarrow }}=0$. \ Using the above results

\begin{equation}
p_{z}=\frac{1}{4}\left\vert \int d\theta \exp \left( i\zeta \right)
\right\vert ^{2},
\end{equation}

where

\begin{equation}
\zeta \left( \theta \right) =-2\int_{0}^{s\left( \theta \right) }ds^{\prime
}\left\vert \mathbf{\kappa }\right\vert .
\end{equation}

\subsection{The case where $\mathbf{\protect\kappa }\left( s\right) $ is
half circle}

Consider the case where $\mathcal{K}=\mathbf{\kappa }\cdot \mathbf{\sigma }$%
, where

\begin{equation}
\mathbf{\kappa }\left( s\right) =\gamma \left( \sqrt{\Lambda ^{2}-\left(
\gamma s\right) ^{2}},0,\gamma s\right) ,  \label{kappa_half_circle}
\end{equation}

$\gamma $ is a non-negative real constant with dimensionality of 1/length, $%
\Lambda $ is a non-negative dimensionless real parameter, and $\left\vert
s\right\vert <\Lambda /\gamma $.

The Zener transition probability is calculated for the case $\Lambda \gtrsim
1$ to lowest order in the adiabatic expansion. \ The following holds

\begin{equation}
\cos \theta =\frac{\gamma s}{\Lambda },
\end{equation}

and $\left\vert \mathbf{\kappa }\right\vert =\gamma \Lambda $, thus

\begin{equation}
\zeta \left( \theta \right) =-2\int_{0}^{s\left( \theta \right) }ds^{\prime
}\left\vert \mathbf{\kappa }\right\vert =-2\Lambda ^{2}\cos \theta ,
\end{equation}

and

\begin{equation}
p_{z}=\frac{1}{4}\left\vert \tint\limits_{-\pi }^{0}d\theta \exp \left(
-2i\Lambda ^{2}\cos \theta \right) \right\vert ^{2}.
\end{equation}

Using the identity

\begin{equation}
\tint\limits_{0}^{\pi }d\theta \exp \left( iz\cos \theta \right) =\pi
J_{0}\left( z\right) ,
\end{equation}

one finds

\begin{equation}
p_{z}\simeq \frac{\pi ^{2}}{4}J_{0}^{2}\left( 2\Lambda ^{2}\right) \text{ \
\ \ \ }\left( \text{for \ }\Lambda \gtrsim 1\right) .  \label{p_z Lambda>1}
\end{equation}

Fig. \ref{half circle num integration} shows an example of numerical
integration of the equation of motion for the case $\Lambda =5$. \ Fig. \ref%
{half circle num integration} (a) shows the 'half circle' $\mathbf{\kappa }%
\left( s\right) $ curve and Fig. \ref{half circle num integration} (b) shows
the evolution of the polarization vector on the Bloch sphere. \ Figure \ref%
{p_z vs. Lambda} shows a numerical calculation of the Zener transition
probability $p_{z}$ as a function of the parameter $\Lambda $. \ As can be
seen in Fig. \ref{p_z vs. Lambda}, $p_{z}$ vanishes for a series of points
we denote as $\Lambda _{n}$ ($n=1,2,3...$). \ The first zero of $p_{z}$ is
at $\Lambda _{1}=1.022$. \ Note however that even though $p_{z}$ vanishes at
the points $\Lambda _{n}$, the evolution becomes truly adiabatic only when $%
\Lambda >>1$.

\FRAME{ftbpFU}{3.4402in}{2.5892in}{0pt}{\Qcb{Example of numerical
integration of the equation of motion for the case $\Lambda =5$. \ On the
left the curve $\mathbf{\protect\kappa }\left( s\right) $ is shown \ and on
the right the evolution of the polarization vector $\mathbf{p}\left(
s\right) $ on the Bloch sphere is seen.}}{\Qlb{half circle num integration}}{%
fig5.gif}{\special{language "Scientific Word";type
"GRAPHIC";maintain-aspect-ratio TRUE;display "USEDEF";valid_file "F";width
3.4402in;height 2.5892in;depth 0pt;original-width 5.6879in;original-height
4.2704in;cropleft "0";croptop "1";cropright "1";cropbottom "0";filename
'Fig5.gif';file-properties "XNPEU";}}

Comparing Eq. \ref{p_z Lambda>1} with the numerical solution seen in Fig. %
\ref{p_z vs. Lambda} shows, as expected, good agreement for $\Lambda \gtrsim
1$. \ For the range $0\leq \Lambda \lesssim 1$, however, we find that the
following can serve as a good approximation

\begin{equation}
p_{z}\simeq J_{0}^{2}\left( \frac{\pi \Lambda ^{2}}{\sqrt{2}}\right) \text{
\ \ \ \ }\left( \text{for \ }0\leq \Lambda \lesssim 1\right) .
\label{T_lambda_small}
\end{equation}

\FRAME{ftbpFU}{3.3883in}{2.7008in}{0pt}{\Qcb{Zener probability $p_{z}$ vs.
the parameter $\Lambda $ calculated numerically.}}{\Qlb{p_z vs. Lambda}}{%
fig6.gif}{\special{language "Scientific Word";type
"GRAPHIC";maintain-aspect-ratio TRUE;display "USEDEF";valid_file "F";width
3.3883in;height 2.7008in;depth 0pt;original-width 4.4789in;original-height
3.5622in;cropleft "0";croptop "1";cropright "1";cropbottom "0";filename
'Fig6.gif';file-properties "XNPEU";}}

\subsection{The case where $\mathbf{\protect\kappa }\left( s\right) $ is a
straight line}

We calculate below $p_{z}$ for the case $\mathcal{K}=\mathbf{\kappa }\cdot 
\mathbf{\sigma }$, where $\mathbf{\kappa }\left( s\right) $ is a straight
line

\begin{equation}
\mathbf{\kappa }\left( s\right) =\Delta \left( 0,1,\gamma s\right) ,
\end{equation}

where $\Delta $ and $\gamma $ are real constants independent of $s$.

For the present case one has

\begin{eqnarray}
\zeta \left( s\right) &=&-2\Delta \int_{0}^{s}ds^{\prime }\sqrt{1+\left(
\gamma s^{\prime }\right) ^{2}} \\
&=&-\frac{\Delta }{\gamma }\left[ \gamma s\sqrt{1+\left( \gamma s\right) ^{2}%
}+\sinh ^{-1}\left( \gamma s\right) \right]  \notag
\end{eqnarray}

and

\begin{eqnarray}
&&-\frac{1}{2}\int_{\pi }^{0}d\theta \exp \left( i\zeta \right) \\
&=&\frac{1}{2}\int_{-\infty }^{\infty }\frac{1}{\cosh z}\exp \left[ -i\frac{%
\Delta }{\gamma }\left( \frac{1}{2}\sinh 2z+z\right) \right] dz.  \notag
\end{eqnarray}

In the limit $\Delta /\gamma \rightarrow \infty $ the phase oscillates
rapidly and consequently $p_{z}\rightarrow 0$. \ The stationary phase points 
$z_{n}$ in the complex plane are found from the condition

\begin{equation}
0=\frac{d}{dz}\left( \frac{1}{2}\sinh 2z+z\right) =\cosh 2z+1,
\end{equation}

thus

\begin{equation}
z_{n}=i\pi \left( n+\frac{1}{2}\right) ,
\end{equation}

where $n$ integer. \ Note, however that the term $1/\cosh z$ has poles at
the same points. \ Using the Cauchy's theorem the path of integration can be
deformed to pass close to the point $z_{-1}=-i\pi /2$. \ Since the pole at $%
z_{-1}$ is a simple one, the principle value of the integral exists. \ To
avoid passing through the pole at $z_{-1}$ a trajectory forming a half
circle "above" the pole with radius $\varepsilon $ is chosen were $%
\varepsilon \rightarrow 0$. \ This section gives the dominant contribution
which is $i\pi R$, where $R$ is the residue at the pole. \ Thus one finds:

\begin{equation}
\left\vert \frac{1}{2}\int_{\pi }^{0}d\theta \exp \left( i\zeta \right)
\right\vert \simeq \exp \left( -\frac{\pi }{2}\frac{\Delta }{\gamma }\right)
.
\end{equation}

The prefactor in front of the exponent is determined by requiring $p_{z}=1$
in the limit $\Delta <<\gamma $, thus

\begin{equation}
p_{z}\simeq \exp \left( -\pi \frac{\Delta }{\gamma }\right) .
\label{p_z staright line}
\end{equation}

\newpage 
\bibliographystyle{plain}
\bibliography{apssamp}

\end{document}